\documentclass[5p]{elsarticle}

\usepackage{amssymb}
\usepackage{amsmath}
\usepackage{booktabs}
\usepackage{lineno}
\journal{Astroparticle Physics}

\newcommand{\cpkky}{{\rm{counts/(keV\,kg\,y)}}}
\newcommand{\DBD}{0$\nu$DBD}

\begin{document}       

\begin{frontmatter}

\title{Discrimination of $\alpha$ and $\beta$/$\gamma$ interactions in a TeO$_2$ bolometer}

\author[LBL]{J.W.~Beeman}
\author[RM1,INFNRM1]{F.~Bellini}
\author[RM1,INFNRM1]{L.~Cardani} 
\author[RM1,INFNRM1]{N.~Casali} 
\author[INFNRM1]{I.~Dafinei}
\author[GE,INFNGE]{ S.~Di~Domizio}
\author[RM1,INFNRM1]{F.~Ferroni}
\author[INFNRM1]{ F.~Orio}
\author[INFNMIB]{ G.~Pessina}
\author[INFNMIB]{ S.~Pirro}
\author[INFNRM1]{ C.~Tomei}
\author[RM1,INFNRM1]{M.~Vignati\corref{cor1}}\ead{marco.vignati@roma1.infn.it}
\address[LBL]{Lawrence Berkeley National Laboratory, Berkeley, California 94720, USA} 
\address[RM1]{Dipartimento di Fisica, Sapienza Universit\`{a} di Roma, Roma I-00185, Italy} 
\address[INFNRM1]{INFN Sezione di Roma, Roma I-00185, Italy} 
\address[GE]{Dipartimento di Fisica, Universit\`{a} degli studi di Genova, Genova I-16146, Italy} 
\address[INFNGE]{INFN Sezione di Genova, Genova I-16146, Italy} 
\address[INFNMIB]{INFN Sezione di Milano Bicocca, Milano I-20126, Italy}
\cortext[cor1]{Corresponding~author}

\begin{abstract}

TeO$_2$ crystals have proven to be  superb bolometers for the search  of neutrinoless double beta decay in many respects. However, if used alone, they do not exhibit any feature 
that allows to discriminate an $\alpha$ energy deposit from a $\beta / \gamma$ one. This fact limits their ability to reject the background due to natural radioactivity
and eventually affects the sensitivity of the search.
In this paper we show the results of a TeO$_2$ crystal where, in coincidence with its bolometric heat signal, also the luminescence light  escaping the crystal is recorded. The results show that we are able to measure the light produced by $\beta / \gamma$ particles, which can be
explained as due to Cerenkov emission. No light is detected from $\alpha$ particles, allowing the rejection of this background source.

\end{abstract}

\begin{keyword}
neutrino mass \sep double beta decay \sep bolometers \sep Cerenkov detector 
\PACS{14.60.Pq \sep 23.40.-s\sep 07.57.Kp \sep 29.40.Ka }
\end{keyword}

\end{frontmatter}

\section{Introduction}
Mysteries about neutrinos are several and of different nature. We know that they are neutral 
particles with an extraordinary little mass compared to the one of all the other particles. 
Although they are massive, we have not succeeded yet in measuring their mass. We do not know 
if the neutrino is a particle different from its antiparticle or rather, as hypothesized by 
Majorana~\cite{majorana}, they are the same particle. Majorana observed that the minimal description of spin 1/2 
particles involves only two degrees of freedom and that such a particle, absolutely neutral, 
coincides with its antiparticle. If the Majorana conjecture holds, then it can be possible to 
observe an extremely 
rare process called 
neutrinoless double beta decay (\DBD)~\cite{DBD}.  In the \DBD\ a nucleus decays into another nucleus emitting 
two electrons and no anti-neutrino, thus violating the total lepton number.  
The decay is mediated by the exchange  between two $\beta$ vertexes of a neutrino,
which controls the decay rate through its mass ($m_{\beta\beta}$).

The signature of this process is a monochromatic line
at the Q-value of the decay in the sum energy spectrum of the two electrons.
The sensitivity of an experiment  to the \DBD\  half-life  goes as:
\begin{equation}
S \propto  a \,\epsilon\, \sqrt{\frac{M t}{B \Delta E}}\;.
\end{equation}
It is clear that increasing the isotopic abundance ($a$) and the efficiency of the signal ($\epsilon$) will end up in a linear gain
on the sensitivity. Detector mass ($M$) and time ($t$), on the other hand, contribute only as the square root, as well as background level ($B$) and energy resolution ($\Delta E$). 
The experimental investigation of this process definitely requires a large amount of DBD emitter, in ultra-low background detectors with 
the capability of selecting reliably the signal from the background. 
The bolometric technique offers several advantages and it has been demonstrated viable by the CUORICINO~\cite{Cuoricino} detector in such a convincing way that 
a 1-ton scale experiment, CUORE~\cite{CUORE}, is now in construction. 
Bolometers are low-temperature-operated particle detectors which provide superior energy resolution,
lower energy thresholds and broader material choice than any other conventional device. 
They can be thought as perfect calorimeters, able to fully thermalize the energy released by 
 a particle. 
Up to now, the choice for bolometers as \DBD\ detectors has fallen on natural TeO$_2$ crystals 
that have very good mechanical and thermal properties together with a very large
content of the candidate isotope $^{130}$Te  (34.2$\%$ isotopic abundance \cite{IA}). 
CUORE will be made of 988 TeO$_2$ bolometers of 750~g each, featuring an energy  
 resolution $\Delta E$ of 5 keV FWHM at the Q-value of the decay, which is around 2527 keV~\cite{Redshaw}.

Bolometer-based  \DBD\ searches require however  extremely low levels of background. 
Even if  the background arising from radioactive contaminants in the bolometers  
themselves is reduced drastically, there is still the problem of the surrounding materials. Surface contamination is 
of particular concern. $\alpha$ particles, arising from radioactive contaminations located 
on the surfaces of the detector or of passive elements facing them, can lose part of their energy 
in a few microns and deposit the rest in a detector. This produces an essentially flat background that affects also the Region
of Interest (ROI) around the \DBD\ Q-value~\cite{Pavan}. The expected background in CUORE is about 0.001~\cpkky\ from $\beta / \gamma$ interactions ($B_{\beta\gamma}$)
and  between 0.01 and 0.04~\cpkky\ from $\alpha$'s ($B_\alpha$). Because of the  $\alpha$ background 
the 1-$\sigma$ sensitivity of CUORE to \DBD, assuming
5~y of data taking, is limited to~\cite{CUORE}:
\begin{equation}
S^{CUORE} \sim 1 - 2 \cdot 10^{26}~{\rm y}\;.
\end{equation}

To cover the entire space of the parameters allowed for the
inverted hierarchy of neutrino masses ($m_{\beta\beta} > 20$~meV)~\cite{DBD}, the sensitivity must be at least $15\cdot 10^{26}$~y, estimate obtained by averaging the Nuclear Matrix Elements of Ref.~\cite{NME}.
 
Although TeO$_2$ crystals are extremely good bolometers, the shape of the
bolometric response does not allow any discrimination of $\alpha$ particle signals  with respect to $\beta / \gamma$'s ones~\cite{LightBol}. 
The natural way to discriminate this background would be  to use a
scintillating bolometer~\cite{Pirro}. 
In such a device the simultaneous and independent read out of the
heat and the scintillation light permits to discriminate events due to $\beta / \gamma$, $\alpha$ and neutrons thanks to
their different scintillation yield.
For what is known to date, at bolometric temperatures (10~mK), TeO$_2$ crystals do not scintillate.
The advantage, however, offered by this material in terms of bolometric performances and natural isotopic abundance are a strong 
motivation to pursue another, even if extremely challenging, option: the read-out of the Cerenkov light emitted  in $\beta / \gamma$ interactions (and not in $\alpha$ ones).
This possibility has been pointed out in Ref.~\cite{Tabarelli} and  this phenomenon may explain what 
has been reported in Ref.~\cite{deMarcillac}. The Cerenkov light emitted in visible wavelengths is predicted to be about 350~eV at the \DBD\ Q--value (2.5 MeV). This value is extremely low compared to usual scintillating bolometers, which emit tens of keV at the same energy.

In this paper we report the observation of light emitted by electrons in a TeO$_2$ bolometer, which allows the
discrimination of $\beta / \gamma$ interactions from $\alpha$ interactions. The amount of light is compatible with the
Cerenkov emission, even though the scintillation hypothesis cannot be discarded.
The outline of this paper is the following: in Section~2 the experimental set-up is described, 
in Section~3 the data analysis is presented,  in Section~4 the results are given and in Section~5 their potential
impact on future experiments is discussed.

\section{Experimental set--up} 

We operated as bolometer a  3.0$\times$2.4$\times$2.8 cm$^3$ TeO$_2$ crystal, whose
weight was 116.65~g.
The crystal was doped with natural samarium, which contains 15\% of $^{147}$Sm, a long living 
isotope (T$_{1/2}$ = 1.07$\times$10$^{11}$~y~\cite{SmHL}) that undergoes $\alpha$ decay 
with a Q--value of 2310$\pm$1 keV~\cite{SmQ}.  This
decay  allowed a direct
analysis of the behavior of $\alpha$'s in an energy region close to the  \DBD.  
The details about the crystal growth and the doping process can be found in Ref.~\cite{Roma TeO2:Sm}.

To detect light, we faced to the TeO$_2$ crystal a high sensitivity dark bolometer~\cite{Pirro2}. 
The light detector (LD) consisted of a 66~mm diameter 1~mm thick pure Ge
crystal, covered with a 600 $\mathring{A}$  layer of SiO$_2$
to ensure good light absorption.  A reflecting foil (3M VM2002)
was placed around the TeO$_2$ crystal to enhance the light collection.

The temperature sensor of the TeO$_2$ crystal was a Neutron Transmutation Doped (NTD) germanium thermistor \cite{NTD}
of $3\times$3$\times$1~mm$^3$, thermally coupled to the crystal surface by means of 9 epoxy glue spots of about 0.6 mm diameter and 50 $\mu$m height.
The LD was equipped with two NTD-Ge thermistors of 3$\times$1.5$\times$0.4 mm$^3$, labeled as L1 and L2.
The TeO$_2$ crystal  and the LD were held in a copper structure by Teflon (PTFE) supports, thermally coupled to the mixing chamber of a dilution refrigerator which kept the system at  a temperature of about 10~mK. 

The detector was operated deep underground 
in the Gran Sasso National Laboratories in Italy, in the CUORE R$\&$D  cryostat.
One of the purposes of the test was to assess the performances of  light detectors for the LUCIFER R$\&$D on
scintillating bolometers~\cite{LUCIFER}.

The read--out of the thermistors was performed via a cold pre--amplifier stage, located inside the cryostat,
and a second amplification stage, located on the top of the cryostat at room temperature. 
After the second stage, the signals were filtered by means of an anti-aliasing 6-pole active Bessel filter (120 db/decade),
and then fed into a NI PXI-6284 analog-to-digital converter operating at a sampling frequency of  2~kHz. 
The Bessel cutoff was set at 120~Hz on L1 and L2, and at 80~Hz on the TeO$_2$.
The details of the cryogenic facility and of the electronics can be found in \cite{HallC cryostat1, HallC cryostat2, HallC cryostat3}.

The trigger was software generated on each bolometer. When it fired, waveforms 2~s long were saved on disk.
If the trigger fired on the TeO$_2$, waveforms from L1 and L2 were anyhow saved, irrespective of their trigger.
To maximize the signal to noise ratio, waveforms were processed offline with the optimum filter algorithm~\cite{OF}.
The main parameters of the bolometers are reported in Tab.~\ref{Table:parameters_crystals}. The rise and decay times
of the pulses are computed as the time difference between the 10\% and the 90\% of the leading edge, 
and the time difference between the 90\% and 30\% of the trailing edge, respectively. 
The intrinsic energy resolution of the detector is evaluated from the fluctuation of the detector noise, after the application of the optimum filter. 
\begin{table}[tbp]
\centering
\caption{Parameters of the bolometers. Amplitude of the signal before  amplification  ($A_S$),  intrinsic energy resolution after the application of the optimum filter  ($\sigma$), rise ($\tau_r$) and decay ($\tau_d$) times of the pulses (see text).}
\begin{tabular}{lcccc}
\toprule
               & $A_S$            &$\sigma$    &$\tau_{r}$  &$\tau_{d}$\\
               &[$\mu$V/MeV]      &[keV RMS]               &[ms]     &[ms]   \\
\midrule
TeO$_2$:Sm     &43                &1.31              &15      &116    \\
\midrule	
Ge (L1)        &1.1$\times$10$^3$   & 0.127          &3      &5\\
\midrule
Ge (L2)        &2.3$\times$10$^3$   & 0.097          &3      &10\\
\bottomrule
\end{tabular}
\label{Table:parameters_crystals}
\end{table}


\section{Data collection and analysis} 

\begin{figure}[t]
\centering
\includegraphics[width=0.48\textwidth,clip=true]{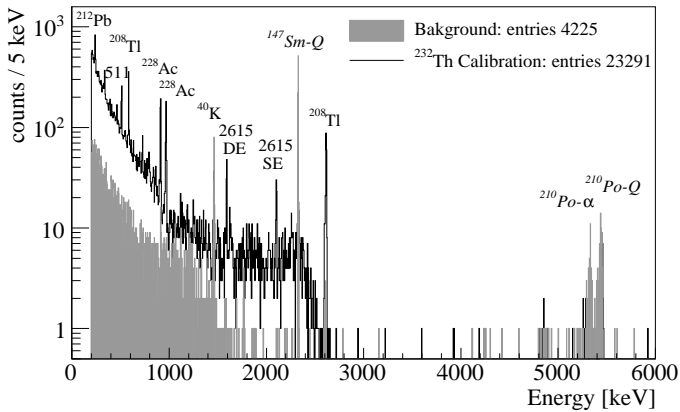}
\caption{Energy spectrum measured with the TeO$_2$ bolometer. See description in the text.}
\label{fig:energy_HD}
\end{figure}

We collected 18.5 days of data.  The detector was exposed to a $^{232}$Th source 
placed outside the cryostat
for 7.7 additional days, to provide statistics to the
$\gamma$ sample, and to perform the energy calibration of the TeO$_2$ bolometer.  
The measured energy spectra with the source (calibration data) and without the source (background data)
are shown in Fig.~\ref{fig:energy_HD}. 
The peak at 2310~keV
corresponds to the $\alpha$ decay of $^{147}$Sm.  The two
peaks  around 5400~keV are due to the $\alpha$ decay of $^{210}$Po, a
contaminant of the TeO$_2$ crystal.  The remaining peaks are $\gamma$'s
from the $^{232}$Th source, except for the peak at 1461~keV, which is a $\gamma$
from $^{40}$K contamination of the cryostat. 
Both the single escape peak (SE) and the double escape peak (DE) of the 
2615~keV $\gamma$ from $^{208}$Tl are visible. The DE is of particular interest
because it is a single site production of $\beta$'s (and  $\beta^+$'s), like the \DBD.   
 
The LD was permanently
exposed to a $^{55}$Fe source, placed on the LD surface opposite to the TeO$_2$ crystal. 
The source produces two X-rays at 5.9 and 6.5~keV, which are  used for the absolute calibration
of the bolometer.

Signals from the TeO$_2$ are very large compared to the noise, which is 1.3 keV
RMS, and the signal amplitude is simply estimated from the maximum of
the filtered signal.  On the LD we expect very small amounts of light
from the TeO$_2$ crystal.  As said previously, the energy released by Cerenkov light in visible wavelengths is predicted to be 350 eV at 2.5 MeV~\cite{Tabarelli}, and the collection  efficiency, not precisely evaluated, is in the range 10-60\%. 
The  signal from the LD is therefore at the level of the noise, which is 127 eV RMS on L1 and 97
eV RMS on L2. 
 We applied the maximum search algorithm to the LD waveforms and we were not able to detect signals.
Irrespective of the energy released in the TeO$_2$, in fact, we measured a constant pedestal at about 350~eV.

To increase the sensitivity of the LD we developed a new method
which leads to a lowering of the energy threshold by about a factor 4 with respect
to the maximum search algorithm~\cite{PipernoVignatiJitter}. The method can be summarized as follows. 
Both TeO$_2$ and LD bolometers respond in milliseconds, 
a time which is orders of magnitude greater than that of the process of light emission. 
The time delay between the two signals depends only on the differences between the thermal and
electronic responses of the two bolometers, and therefore is fixed.
Instead of looking for a maximum, we take the value of the LD filtered
waveform at a fixed time delay with respect to the signal found on
the TeO$_2$.  To estimate the time delay, we selected events
generated by particles traversing at the same time the LD and the TeO$_2$.
In these events the signals are greater than 200~keV on both bolometers,
allowing an easy identification of the pulses.

\section{Results} 
\begin{figure}[h!]
\centering
\includegraphics[width=0.47\textwidth,clip=true]{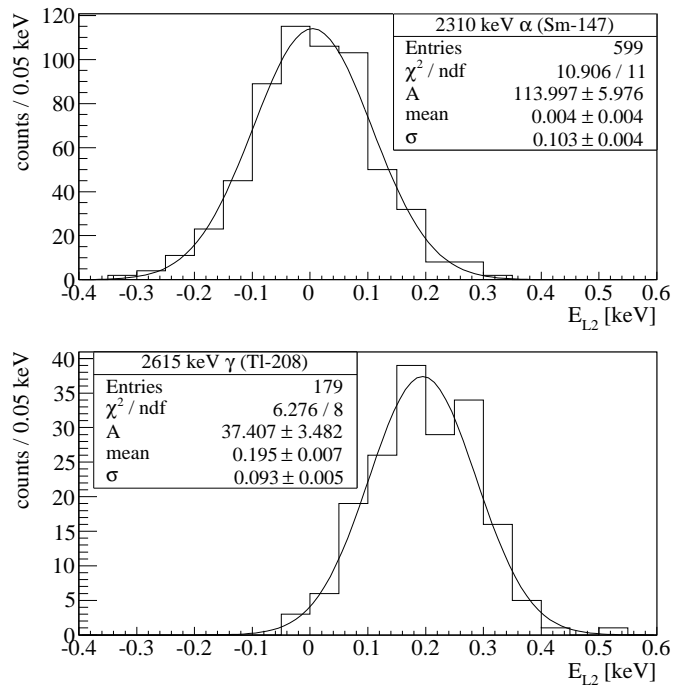}
\caption{Distribution of the light emitted from  2310~keV  $\alpha$ decays (top) and 2615~keV $\gamma$-ray interactions (bottom),
fitted with a Gaussian function $A\cdot\exp[-(E_{L2}-{\rm mean})^2/2\sigma^2$].
Data  from thermistor L2. }
\label{fig:L2_light}
\end{figure}

We use the calibration data to estimate the light emitted from $\gamma$ interactions, and the background
data to estimate the light from $\alpha$ interactions. 
The distributions of the light measured using thermistor L2 for the two peaks close to the 0$\nu$DBD energy, the
$\gamma$-ray at 2615~keV and the $\alpha$ decay at 2310~keV, are shown in Fig.~\ref{fig:L2_light}.
The average energy  is $195\pm7$~eV for the $\gamma$-ray and $4\pm4$~eV
for the $\alpha$ decay, and the resolutions are compatible with the intrinsic resolutions reported in Tab.~\ref{Table:parameters_crystals}.

\begin{figure}[tbp]
\centering
\includegraphics[width=0.235\textwidth,clip=true]{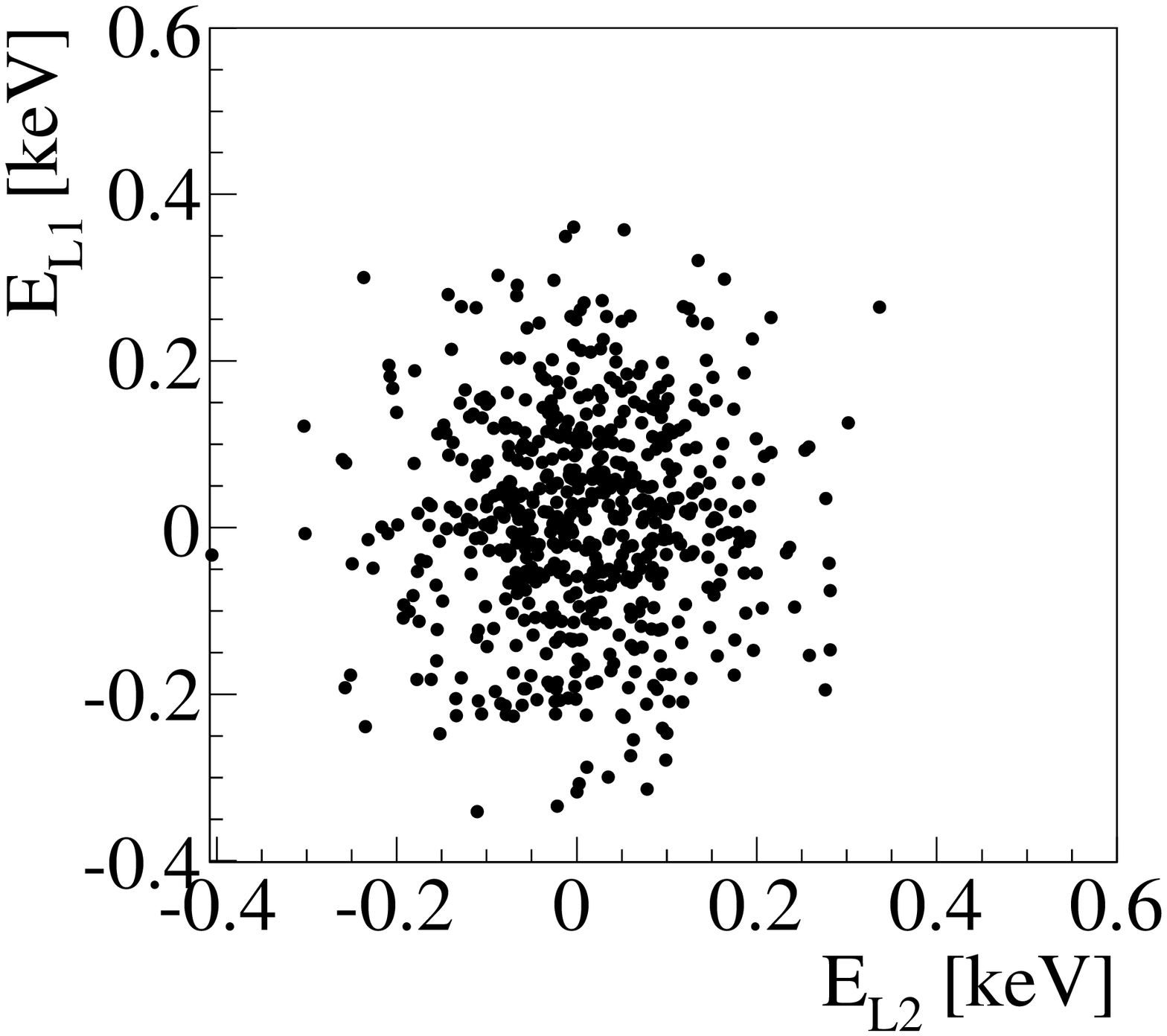}
\includegraphics[width=0.235\textwidth,clip=true]{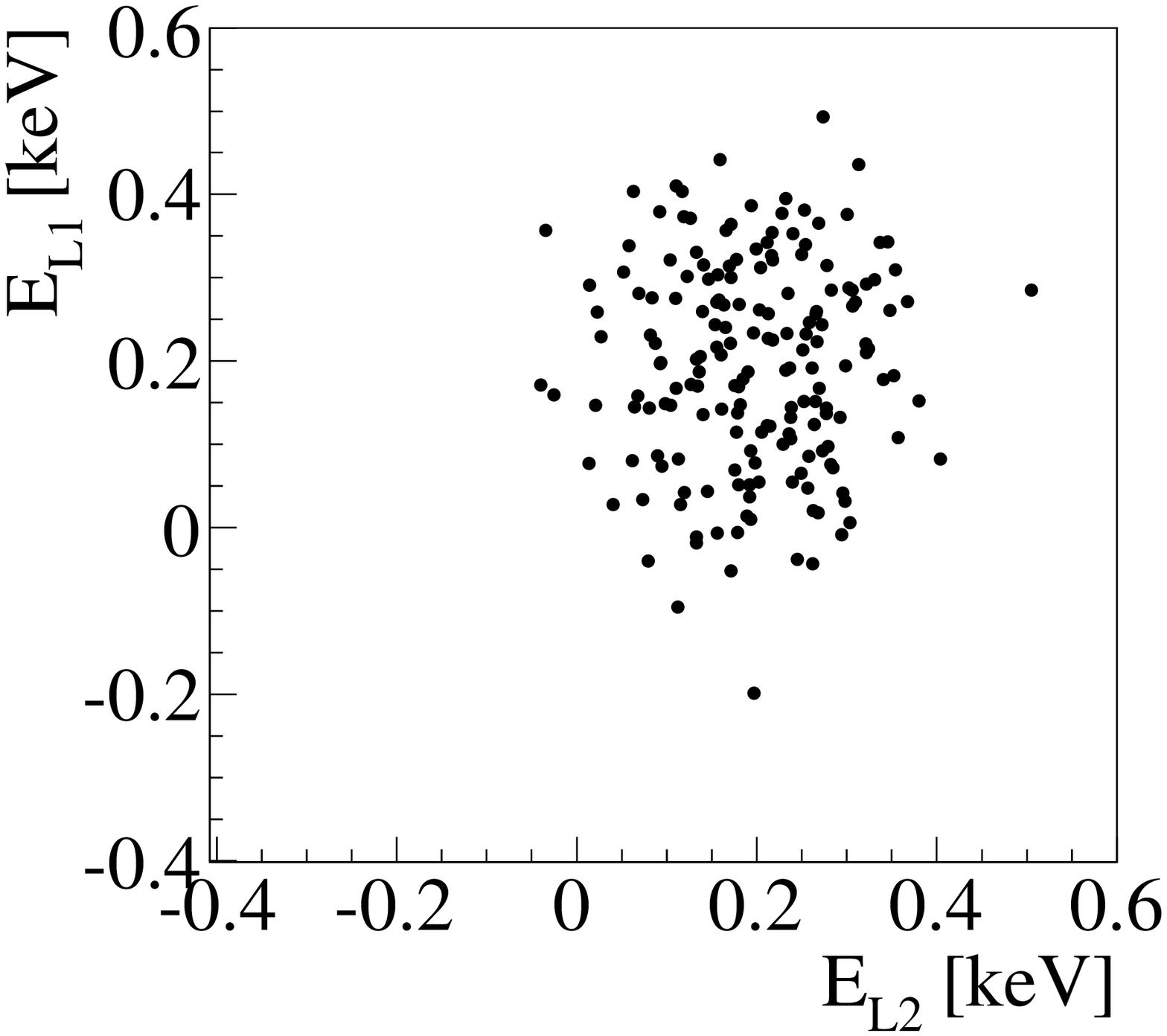}
\caption{Light detected using thermistor L1 versus light detected using thermistor L2 on  2310~keV  $\alpha$ decays (left)
and  2615~keV $\gamma$-ray interactions (right). The correlation is negligible in both cases, being 4\% for $\alpha$ decays
and 3\% for $\gamma$-rays.}
\label{fig:L1_vs_L2}
\end{figure}

\begin{figure}[tbp]
\centering
\includegraphics[width=0.48\textwidth,clip=true]{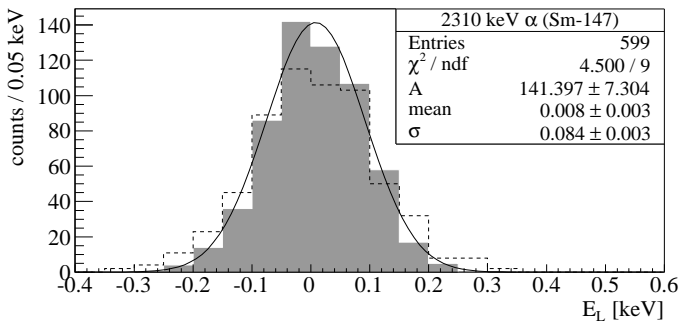}
\caption{Distribution of the light emitted from  2310~keV  $\alpha$ decays.  Combination of the energies
from the two thermistors (solid) and energy  from  thermistor L2 alone (dashed). }
\label{fig:L1L2_Sm147}
\end{figure}

The results obtained using  thermistor L1 in place of L2 are similar, even though the energy resolution of L1
is worse by approximately  30\%.
The correlation between the energies measured by the two thermistors is negligible (Fig.~\ref{fig:L1_vs_L2}), implying 
that the resolution is dominated by the noise from some component of the read-out system and not by 
the germanium slab.
To improve the energy resolution, we combine the energies measured by the two thermistors into an unique energy estimator: 
\begin{equation}
E_L  = \frac{w_1 E_{L1} + w_2 E_{L2}}{w_1+w_2}\;,
\label{eq:Lcomb}
\end{equation}
where $w_{1,2} =  1/\sigma_{L1,L2}^2$.
The resolution on the $\alpha$ peak improves from $103\pm4$ eV to $84\pm3$ eV (Fig.~\ref{fig:L1L2_Sm147}).

\begin{figure}[tbp]
\centering
\includegraphics[width=0.48\textwidth,clip=true]{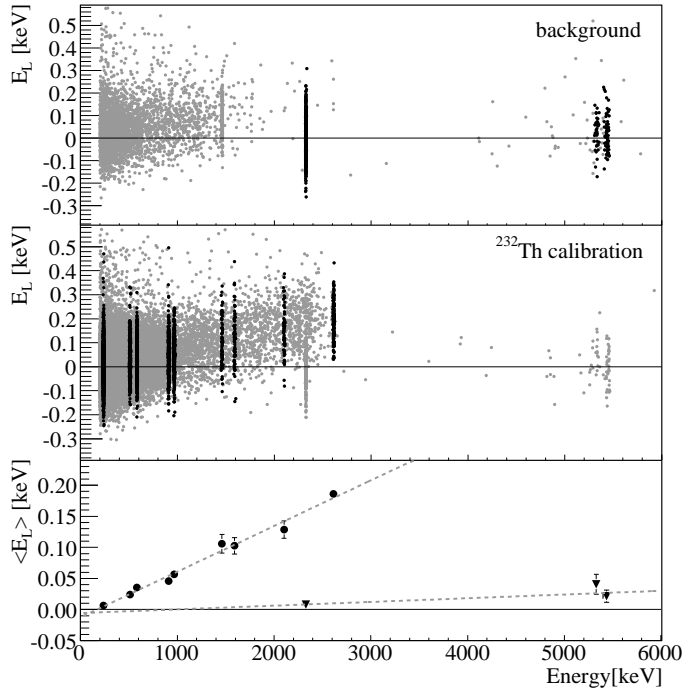}
\caption{Distribution of the emitted light $E_L$ versus particle energy.
In the top (background data) and middle (calibration data)  figures  the peaks labeled in Fig.~\protect\ref{fig:energy_HD} are marked in black. In the bottom figure
the average light for each peak is shown, using  background data for the $\alpha$ peaks (triangles) and
 calibration data for the $\gamma$ peaks (circles). $\gamma$ and $\alpha$ points are fitted with two different lines.}
\label{fig:LvsHeat}
\end{figure}

The distribution of $E_L$ versus particle energy is shown 
in Fig.~\ref{fig:LvsHeat}. The peaks in the background data (top) and in the calibration data
(middle)  are used to estimate the light emitted by $\gamma$ interactions and 
$\alpha$ decays, respectively. 
We estimate the average light of each peak, $<E_L>$, by means of Gaussian fits, fixing the
variance of the Gaussian to the value estimated on the $^{147}$Sm peak, and
we perform separate linear fits for  $\alpha$'s and $\gamma$'s to the distribution of $<E_L>$  versus energy  (bottom of the figure): 
\begin{equation}
<E_{L}> = a + b\cdot  {\rm Energy}\;.
\label{eq:lightdep}
\end{equation}
The results of the linear fits are reported in Tab.~\ref{tab:fit_l_vs_h}.
The  value of $b_\alpha$ is compatible with zero within two standard deviations, which is expected
if the light we detect from $\gamma$ interactions is due to Cerenkov radiation. 
The line interpolating the $\gamma$'s indicates that the 2615 keV DE peak, which is actually due to $\beta^-\beta^+$ interactions, behaves like $\gamma$ peaks.
From the line parameters we evaluate the threshold on the particle energy, 
$E^{th} = -a_\gamma / b_\gamma$, to be  $171\pm23$ keV.
\begin{table}[tbp]
\centering
\caption{Parameters of the linear fits to the $\gamma$ and $\alpha$ distributions in Fig.~\protect\ref{fig:LvsHeat} (bottom).
Intercept ($a$), slope ($b$), correlation between $a$ and $b$ ($\rho_{ab}$), and $\chi^2$/ndf.}
\begin{tabular}{ccccc}
\toprule
                               & $a$ [eV] & $b$ [eV/ MeV]  & $\rho_{ab}$ [\%] & $\chi^2$/ndf  \\ 
\midrule
$\alpha$   &            $\phantom{1}-6\pm9$      & $\phantom{7}6\pm3$  & -93 & 1/1\\
\midrule
$\gamma$               &  $-13 \pm 2$  &  $73\pm2$   & -76 & 11/7\\

\bottomrule
\end{tabular}
\label{tab:fit_l_vs_h}
\end{table}

Although the energy threshold for Cerenkov emission by
an electron in TeO$_2$ is about 50 keV, the value we measure is compatible  with  the fact
that in $\gamma$ interactions multiple electrons are produced.
The scintillation hypothesis, however, cannot be ruled out completely, since we cannot measure
the wavelenght spectrum, nor the time characteristics of the light signal.  Moreover the light from $\alpha$'s  (if any)
could be even smaller than what we measure, since cross talks in the electronics cannot be excluded at this level of precision.

\section{Perspectives} 

We study the improvements that can be obtained with an experiment
like CUORE, but equipped with light detectors.
The light emitted from $\alpha$'s is considered null independently
of the particle energy.  The light emitted from $\beta / \gamma$'s follows
Eq.~\ref{eq:lightdep}, which implies for a $\beta / \gamma$ with 0$\nu$DBD energy
$<E_L^{0\nu}> = 173$~eV.  
By applying a threshold $T$ on the detected light ($E_L>T$), the
$\alpha$ background is reduced by a factor $\epsilon_{\alpha}$. 
The threshold affects also the \DBD\ signal and the $\beta / \gamma$ background,
reducing them by a different factor, $\epsilon_{\beta\gamma}$. The sensitivity
of the experiment is then modified as follows:
\begin{equation}
S(T) = S^{CUORE}\epsilon_{\beta\gamma}(T)  \sqrt{\frac{ B_\alpha + B_{\beta\gamma}   }{\epsilon_\alpha(T)  B_\alpha + \epsilon_{\beta\gamma}(T)  B_{\beta\gamma}   }}\;.
\end{equation}
The value of $T$ which maximizes $S$, assuming  $B_\alpha = 0.01$ 
and $B_{\beta\gamma}= 0.001~\cpkky$, is found to be 148~eV, corresponding to $S^{\rm max} = 4.0\cdot10^{26}$~y.

\begin{figure}[t]
\centering
\includegraphics[width=0.45\textwidth,clip=true]{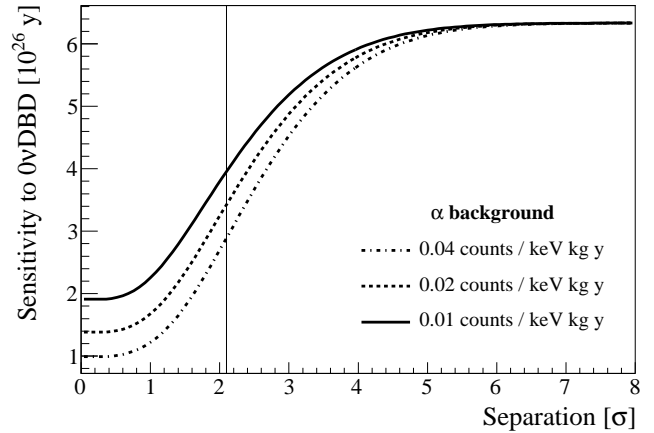}
\caption{1-$\sigma$ sensitivity to the 0$\nu$DBD half-life as a function of the separation between $\beta / \gamma$'s and $\alpha$'s. 
The   $\alpha$ background is within 0.04 and 0.01~\cpkky\ and we make the hypothesis
that  the $\beta / \gamma$ background is 0.001~\cpkky\ . The sensitivity of CUORE corresponds to zero separation, while the separation achieved in this work is 2.1~$\sigma$. }
\label{fig:sensitivity}
\end{figure}

We define the separation of the 0$\nu$DBD from $\alpha$'s as the distance of  $<E_L^{0\nu}>$ from zero energy in units of the energy resolution.  In Fig.~\ref{fig:sensitivity}
we show  $S^{\rm max}$ as a function of the separation for an experiment equal to CUORE,
but equipped with light detectors, for different values of the $\alpha$ background. The
separation obtained in this work is $173/84 = 2.1~\sigma$.
The  sensitivity plateau is reached when the separation is greater than $4-5~\sigma$, i.e. about a factor  2 higher than the present work.
 The sensitivity would then be from 3 to 6 times higher than in CUORE, depending
on the amount $\alpha$ background that will be present.

The separation can be improved by reducing the noise, which seems
dominated by the read-out of the bolometer (thermistor, cables
or electronics).  It can also be improved by increasing the 
efficiency of light collection. If the Cerenkov hypothesis
is confirmed,  in fact, about the $45\%$ of the light  is emitted in the UV range, a region where the reflectivity of the foil
we used (VM2002 from 3M) is less than $30\%$. Wide band reflectors, like PTFE and alluminium, should substantially increase
the amount of light detected.

If this technique could be complemented with a 95$\%$ enrichment of the crystals  in $^{130}$Te, 
an additional factor 3 could be gained, projecting the final sensitivity to the desired value of $15\cdot 10^{26}$~y, capable of probing $m_{\beta \beta}$ as low as 20 meV.

\section{Acknowledgments}
The project LUCIFER has received funding from the European Research Council
under the European UnionÕs Seventh Framework Programme (FP7/2007-2013) /
ERC grant agreement n. 247115.


\begin{thebibliography}{100}

\bibitem{majorana}
E. Majorana,  Il Nuovo Cimento  {\bf 14} 171 (1937).

\bibitem{DBD}
F. T. Avignone, S. R. Elliott and J. Engel, Rev. Mod. Phys. {\bf 80} 481 (2008).

\bibitem{Cuoricino} 
 E. Andreotti  et al.,  Astropart. Phys. {\bf 34} 822 (2011).

\bibitem{CUORE} 
R. Ardito  et al.  arXiv:hep-ex/0501010  (2005);\\
C. Arnaboldi et.al,  Nucl. Instrum. Meth. A {\bf 518} 775 (2004).

\bibitem{IA}
M. A. Fehr, M. Rehkamper and  A. N. Halliday, Int. J. Mass Spectrom. {\bf 232} 83 (2004).
  
\bibitem{Redshaw} M. Redshaw, B. J. Mount, E. G. Myers and  F. T. Avignone, Phys. Rev. Lett. {\bf 102} 12502 (2009). 

\bibitem{Pavan}
M. Pavan et al., Eur. Phys. J. A {\bf 36} 159 (2008).

\bibitem{NME} J. Barea and F. Iachello, Phys. Rev. C {\bf 79} 044301 (2009);\\
F. Simkovic et al., Phys. Rev. C {\bf 77} 045503 (2008);\\
O. Civitarese and J. Suhonen, J. Phys. Conf. Ser. {\bf 173}  012012 (2009). 
 
\bibitem{LightBol}
C. Arnaboldi et al. Astropart. Phys. {\bf 34} 797 (2011).

\bibitem{Pirro}
S. Pirro et al. Phys. Atom. Nucl. {\bf 69} 2109 (2006). 

\bibitem{Tabarelli}
T. Tabarelli de Fatis,
Eur. Phys. J. C {\bf 65} 359 (2010).

\bibitem{deMarcillac}
N. Coron et al.,  
Nucl. Instrum. Meth. A {\bf 520} 159 (2004). 


\bibitem{SmHL}
K Kossert, G. J\"org, O. N\"ahle and C. L. V. Gostomski, Appl. Radiat. Isotopes {\bf 67}  1702 (2009). 

\bibitem{SmQ}
M. C. Gupta and R. D. MacFarlane, 
J. Inorg.\ Nucl.\ Chem.\ {\bf 32}  3425 (1970).

\bibitem{Roma TeO2:Sm}
F. Bellini et al., JINST {\bf 5} P12005 (2010).

\bibitem{Pirro2}
S. Pirro et al., Nucl. Instrum.  Meth. A {\bf 559} 361 (2006).
 
\bibitem{NTD}
E. E. Haller et al., 
in Neutron transmutation doping of semiconducting materials, edited by R.D. Larrabee (Plenum Press, New York, 1984), p.~21.

\bibitem{LUCIFER}
F. Ferroni,  J. Phys. Conf. Ser. {\bf 293} 012005 (2011).

\bibitem{HallC cryostat1}
S. Pirro, Nucl. Instrum. Meth. A {\bf 559} 672 (2006).

\bibitem{HallC cryostat2}
C. Arnaboldi, G. Pessina and  S. Pirro, Nucl. Instrum. Meth. A {\bf 559} 826 (2006).

\bibitem{HallC cryostat3} 
C. Arnaboldi et al., Nucl. Instrum. Meth. A {\bf 520} 578 (2004).

\bibitem{OF} E.~Gatti and P.F.~Manfredi, Riv. Nuovo Cimento {\bf 9} 1 (1986).




\bibitem{PipernoVignatiJitter} G.~Piperno, S.~Pirro and M.~Vignati, 
A data analysis method to lower the energy threshold of light detectors coupled to scintillating bolometers, {\it paper in preparation}.

\end{thebibliography}
\end {document}